\documentclass[onecolumn,superscriptaddress,amsmath,amssymb,%
nofootinbib]{revtex4}

\usepackage{graphicx}

\begin{document}

\title{Analytic self-gravitating $4$-Baryons, traversable NUT-AdS wormholes,
flat space-time multi-Skyrmions at finite volume and a novel transition in
the $SU(3)$-Skyrme model}
\thanks{We dedicate this work to Murray Gell-Mann, an eight-leagues giant of
physics.}

\author{Eloy Ay\'on-Beato}
\email{ayon-beato-at-fis.cinvestav.mx} \affiliation{Departamento de
F\'{\i}sica, CINVESTAV--IPN, Apdo.\ Postal 14--740, 07000, CDMX, M\'exico}

\author{Fabrizio Canfora}
\email{canfora-at-cecs.cl} \affiliation{Centro de Estudios Cient\'{\i}ficos
(CECs), Casilla 1469, Valdivia, Chile}

\author{Marcela Lagos}
\email{marcelagosf-at-gmail.com} \affiliation{Instituto de Ciencias F\'isicas
y Matem\'aticas, Universidad Austral de Chile, Valdivia, Chile}

\author{Julio Oliva}
\email{julioolivazapata-at-gmail.com} \affiliation{Departamento de F\'isica,
Universidad de Concepci\'on, Casilla 160-C, Concepci\'on, Chile}

\author{Aldo Vera}
\email{avera-at-cecs.cl} \affiliation{Centro de Estudios Cient\'{\i}ficos
(CECs), Casilla 1469, Valdivia, Chile} \affiliation{Instituto de Ciencias
F\'isicas y Matem\'aticas, Universidad Austral de Chile, Valdivia, Chile}

\begin{abstract}
We construct the first analytic self-gravitating Skyrmions with higher
Baryon charge in four dimensions for the $SU(3)$-Skyrme-Einstein-$\Lambda$
theory by combining the generalized hedgehog ansatz with the approach
developed by Balachandran et al.\ to describe the first (numerical) example
of a non-embedded solution. These are genuine $SU(3)$ analytic solutions
instead of trivial embeddings of $SU(2)$ into $SU(3)$ and its geometry is
that of a Bianchi IX Universe. The Skyrme ansatz is chosen in such a way
that the Skyrme field equations are identically satisfied in the sector
with Baryon charge $4$. The field equations reduce to a dynamical system
for the three Bianchi IX scale factors. Particular solutions are explicitly
analyzed. Traversable wormholes with NUT-AdS asymptotics supported by a
topologically non-trivial $SU(3)$-sigma soliton are also constructed. The
self-gravitating solutions admit also a suitable flat limit giving rise to
Skyrmions of charge 4 confined in a box of finite volume maintaining the
integrability of the $SU(3)$ Skyrme field equations. This formalism
discloses a novel transition at finite Baryon density arising from the
competition between embedded and non-embedded solutions in which the
non-embedded solutions prevail at high density while are suppressed at low
densities.
\end{abstract}

\maketitle

\section{Introduction}

One of the most important results in low-energy quantum chromodynamics (QCD)
is that its effective action becomes the Skyrme model \cite{witten0}; a
Bosonic action for a $SU(N)$-valued scalar field \cite{skyrme}, the physical
case being $SU(3)$. Its solitons, dubbed \textit{Skyrmions}, represent
Fermionic states whose topological charge is the Baryon number
\cite{witten0,bala0}, see also \cite{All,ANW,manton}. These results have been
also generalized to curved space-times \cite{curved1f}. The Skyrme model has
been deeply analyzed not only by its emergence in low energy QCD but also due
to its relevance in General Relativity. For instance, black holes with a
non-trivial Skyrme hair were found using numerical tools in
\cite{lucock,droz} providing the first genuine counterexample to the
well-known ``no-hair'' conjecture; unlike other supposedly ``hairy'' black
holes that were unstable at the end \cite{Bizon:1994dh}, see also
\cite{numerical1,numerical2}. Cosmological applications of the Skyrme model
have also been considered \cite{cosmo,cosmo2}.

Many of the important results in both the Skyrme model and the
Einstein-Skyrme system have been derived numerically due to the highly
non-linear character of the field equations and, until very recently, there
were no analytic solutions with Baryon charge in these models. Analytic
results are not just of academic interest. For instance, it was known that
large isospin chemical potentials lead to Skyrmion instabilities on flat
space. However, it is just lately that this critical behavior in the chemical
potential has been fully understood thanks to an analytic formula
\cite{canfora9,canfora8,Canfora:2018clt} \cite{crist1} \cite{crist2}.
Although the most relevant case corresponds to the $SU(3)$ group, many of the
theoretical and numerical works have been performed for the $SU(2)$ case,
which is already quite difficult in itself but not as much as the former one
that not only exhibits five more generators infinitesimally, but also a
non-constant curvature group manifold. However, since genuine features of the
$SU(3)$ Skyrme model could have very important physical consequences it is
important to pursue any hint contributing to its understanding. In the
seminal works \cite{bala0,Bala1} the first numerical example of a
\textit{non-embedded} solution was constructed. This is a genuine feature of
the $SU(3)$ Skyrme model since the solution has spherical symmetry and at the
same time Baryon charge equal to $2$; in contrast to the $SU(2)$ Skyrme model
where the only stable solution with spherical symmetry, in the hedgehog
sense, has unit Baryon charge. These pioneering ideas have been generalized
in \cite{kopeilovich,ioannidou1,ioannidou2,ioannidou3}. Remarkably enough, in
the present paper we provide the first analytic examples of these types. Very
recently, following the techniques developed in
\cite{canfora2,yang1,ferreira,canfora10,Giacomini:2017xno,Astorino:2018dtr},
the first analytic self-gravitating $SU(2)$ Skyrmions have been constructed
in \cite{canfora6}. Here it will be shown that the configurations found in
\cite{canfora6} can be generalized to genuine self-gravitating and
topologically non-trivial $SU(3)$ configurations. The resulting ansatz is
based on the $SO(3)$ subgroup of $SU(3)$ that has been introduced in
\cite{bala0,Bala1}. Moreover, we can define a suitable flat limit of these
self-gravitating configurations in which they remain at the same time
topologically non-trivial and integrable. Such a limit corresponds to
confining the $4$-Baryon (our analytic configuration with topological charge
equals to $4$) in a finite volume. This construction discloses a novel finite
density transition: the non-embedded solutions prevail at high Baryon density
while the trivially-embedded solutions prevail at low Baryon density.

The paper is organized as follows: in the second section, the Einstein-Skyrme
model is introduced. In the third section, the gravitating $4$-Baryon is
constructed. In the fourth section, the NUT-AdS wormhole is described. In the
fifth section, $4$-Baryons living within a finite flat spatial volume are
introduced. In the sixth section, the possibility of a phase transition
arising from the competition between embedded and non-embedded solutions is
discussed. Eventually, some conclusions and perspectives are presented.

\section{The Einstein-Skyrme theory}

The Einstein-Skyrme system in presence of a cosmological constant is
described by the action
\begin{equation}
I[g,U]=\int d^{4}x\sqrt{-g}\left( \frac{R-2\Lambda }{2\kappa }+\frac{K}{4}
\mathrm{Tr}[A^{\mu }A_{\mu }+\frac{\lambda }{8}F_{\mu \nu }F^{\mu \nu
}]\right) ,  \label{skyrmeaction}
\end{equation}
where $R$ is the Ricci scalar, $A_{\mu }=U^{-1}\nabla _{\mu }U$, with $U\in
SU(N)$ and $\nabla _{\mu }$ the covariant derivative. Moreover $F_{\mu \nu
}=[A_{\mu },A_{\nu }]$, $\Lambda $ is the cosmological constant, $\kappa $
the gravitational constant, and the positive couplings $K$ and $\lambda $ are
fixed by experimental data. In our convention $c=\hbar =1$ and Greek indices
run over the four dimensional space-time with mostly plus signature.

The complete Einstein-Skyrme field equations read
\begin{equation}
\nabla ^{\mu }A_{\mu }+\frac{\lambda }{4}\nabla ^{\mu }[A^{\nu },F_{\mu \nu
}]=0\ ,\quad G_{\mu \nu }+\Lambda g_{\mu \nu }=\kappa T_{\mu \nu }\ ,
\label{skyrme-einstein}
\end{equation}
where $G_{\mu \nu }$ is the Einstein tensor, and the Skyrme energy-momentum
tensor is defined by
\begin{align}
T_{\mu \nu } = & -\frac{K}{2}\mathrm{Tr}\left[ A_{\mu }A_{\nu }
-\frac{1}{2}g_{\mu \nu }A^{\alpha }A_{\alpha }
+\frac{\lambda }{4}\biggl( g^{\alpha \beta
}F_{\mu \alpha }F_{\nu \beta }
-\frac{1}{4}g_{\mu \nu }F_{\alpha \beta }F^{\alpha \beta}\biggl) \right] .
\label{tmunu}
\end{align}
The winding number for a given solution is given by
\begin{equation}
B=\frac{1}{24\pi ^{2}}\int \rho _{\text{B}}\ d^3x \ ,\quad
\rho _{\text{B}}=\text{Tr}[\epsilon
^{ijk}A_{i}A_{j}A_{k}]\ .  \label{winding}
\end{equation}
When the topological density $\rho _{\text{B}}$ is integrated on a space-like
surface, $B$ represents the Baryon number of the configuration.

\section{$SU(3)$ self-gravitating $4$-Baryon}

Before presenting the new results, first we will shortly describe the trivial
embedding into $SU(3)$ of the $SU(2)$ self-gravitating solution found in
\cite{canfora6}.

\subsection{The $SU(2)$ embedded self-gravitating Skyrmion}

The $SU(2)$ generalized hedgehog ansatz reads
\begin{gather} \notag
U(x^{\mu })=Y^{0}(x^{\mu })I\pm Y^{i}(x^{\mu })\Lambda _{i} \ , \quad \left(
Y^{0}\right) ^{2}+Y^{i}Y_{i}=1 \ , \\
\Lambda _{i}=(\lambda _{1},\lambda _{2},\lambda _{3}) \ , \quad
Y^{0}=\cos {\alpha }, \quad Y^{i}=n^{i}\sin {\alpha } \ ,  \label{sessea1}
\end{gather}
where $\ n^{1}=\sin{\Theta }\cos{\Phi},\ n^{2}=\sin{\Theta }\sin {\Phi },\
n^{3}=\cos {\Theta }\,$ and $\left\{ \lambda _{j}\right\} _{j=1,..,8}$ are
the Gell-Mann matrices. Notice that $\{\lambda _{1},\lambda _{2},\lambda
_{3}\}$ generate the $SU(2)$ subgroup of $SU(3)$. The above scalar functions
are chosen in \cite{canfora6} as
\begin{equation}
\Phi =\frac{\gamma +\varphi }{2}\ ,\quad \tan \Theta
=\frac{\cot \left( \frac{\theta }{2}\right) }
{\cos \left( \frac{\gamma -\varphi }{2}\right) }\ ,\quad
\tan \alpha =\frac{\sqrt{1+\tan ^{2}\Theta }}{\tan \left( \frac{\gamma
-\varphi }{2}\right) }\ ,  \label{sessea2}
\end{equation}
while the metric is
\begin{equation}
ds^{2}=-dt^{2}+\rho \left( t\right) ^{2}\left[ (d\gamma +\cos \theta
d\varphi )^{2}+d\theta ^{2}+\sin ^{2}\theta d\varphi ^{2}\right] \ ,
\label{simplem}
\end{equation}
with the range of coordinates
\begin{equation}
0\leq \gamma <4\pi \ ,\;\;\;\ 0\leq \theta <\pi \ ,\;\;\;\ 0\leq \varphi
<2\pi \ ,  \label{range1}
\end{equation}
uniquely fixed by requiring the regularity of the metric. With this ansatz
one can verify that the Skyrme field equations are identically satisfied,
while the Einstein equations reduce to \cite{canfora6}
\begin{equation}
\dot{\rho}^{2}=\frac{\Lambda }{3}\rho ^{2}+\frac{\lambda \kappa K}{32\rho
^{2}}+\frac{\kappa K-2}{8}\ ,\qquad \ddot{\rho}=\frac{\Lambda }{3}\rho
-\frac{\lambda \kappa K}{32\rho ^{3}}\ .  \label{isoskSU(2)}
\end{equation}
Using Eqs.~(\ref{winding}) and (\ref{range1}) the Baryon number of this
configuration turns out to be $B=1$.

\subsection{The new $SU(3)$ non-embedded self-gravitating $4$-Baryon}

To construct a self-gravitating solution with higher Baryonic charge we use
the remarkable ansatz introduced in \cite{bala0,Bala1} for a diBaryon in flat
space-time. This ansatz is constructed with the subalgebra of the Gell-Mann
matrices generating the subgroup $SO(3)\subseteq SU(3)$, namely $\{\lambda
_{7},-\lambda _{5},\lambda _{2}\}$. The $U$ field reads
\begin{align}
U_{\text{B}} =\exp \left( i\psi \right) \mathbf{1}_{3\times 3}
+i\sin \left( \chi \right) \exp \left( -\frac{i\psi }{2}\right) \mathbf{T}
+\left( \cos \left( \chi \right) \exp \left( -\frac{i\psi }{2}\right)
-\exp \left( i\psi \right) \right) \mathbf{T}^{2} \ ,  \label{bala}
\end{align}
\begin{equation}
\mathbf{T}=\vec{\Lambda}\cdot \hat{n} \ , \quad
\hat{n}=(\sin \Theta \cos \Phi ,\sin \Theta \sin \Phi ,\cos \Theta ) \ , \quad
\vec{\Lambda}=(\lambda _{7},-\lambda_{5},\lambda _{2}) \ .  \label{bala2}
\end{equation}
In Refs.~\cite{bala0,Bala1} $\psi $ and $\chi $ are the radial profiles of
the spherically symmetric $SU(3)$\ Skyrmion, while $\Theta $ and $\Phi $ are
chosen as the spherical angles. Here, we promote the four functions to new
profiles depending on the coordinates of a generally curved space-time, which
must be determined by solving the Skyrme field equations.

In the flat case, the Baryon charge can be different from zero even if the
profile $\psi $ vanishes \cite{bala0,Bala1} but the field equations require
$\psi$ to be non-trivial. On the other hand, in the Einstein-Skyrme case one
can take $\psi=0$ for the curved space-times we consider as it will be now
discussed.

At this point, we need an educated ansatz for the four functions appearing in
the above ansatz. A slight modification of the generalized hedgehog ansatz of
\cite{canfora6} defined by $\psi =0$ and
\begin{equation}
\Phi =\frac{\gamma +\varphi }{2}\ ,\quad \tan \Theta =\frac{\cot \left( \frac{
\theta }{2}\right) }{\cos \left( \frac{\gamma -\varphi }{2}\right) }\ ,\quad
\tan \left( \frac{\chi }{2}\right) =\frac{\sqrt{1+\tan ^{2}\Theta }}{\tan
\left( \frac{\gamma -\varphi }{2}\right) }\ ,  \label{skbala}
\end{equation}
does the job. Indeed, the $SU(3)$ Skyrme configuration
(\ref{bala})-(\ref{skbala}) identically satisfies the Skyrme field equations
on any metric of the form (\ref{simplem})-(\ref{range1}).

Thus, Einstein equations with the energy-momentum tensor in Eq.~(\ref{tmunu})
corresponding to the Skyrme configurations defined in
Eqs.~(\ref{bala})-(\ref{skbala}) reduce to
\begin{equation}
\dot{\rho}^{2}=\frac{\Lambda }{3}\rho ^{2}+\frac{\lambda \kappa K}{8\rho ^{2}
}+\frac{2\kappa K-1}{4}\ ,\qquad \ddot{\rho}=\frac{\Lambda }{3}\rho -\frac{
\lambda \kappa K}{8\rho ^{3}}\ .  \label{isosk}
\end{equation}
One can write down the most general solution of Eqs.~(\ref{isosk}) following
the analysis of \cite{canfora10}. The above Eq.~(\ref{isosk}) in the $SU(3)$
case are very similar to the $SU(2)$ one in Eq. (\ref{isoskSU(2)}), but one
can notice that the coefficients appearing are different; in fact, the
$SU(3)$ energy-momentum contribution is four times that of $SU(2)$. This
difference also appears when computing the topological charge density $\rho
_{\text{B}}^{\text{SU(3)}}$ of the $SU(3)$ ansatz, defined in
Eqs.~(\ref{bala})-(\ref{skbala}), and the corresponding $\rho
_{\text{B}}^{\text{SU(2)}}$ of the $SU(2)$ ansatz, defined in
Eqs.~(\ref{sessea1}) and (\ref{sessea2}),
\begin{equation*}
\rho _{\text{B}}^{\text{SU(3)}}=6\sin {\theta }\ ,\qquad
\rho _{\text{B}}^{\text{SU(2)}}=\frac{3}{2}\sin {\theta }\ .
\end{equation*}
Taking into account Eqs.~(\ref{winding}) and (\ref{range1}), the topological
charge of the $SU(3)$ configuration is $B=4$. These are genuine $SU(3)$
configurations: in $SU(2)$ one cannot obtain configurations with topological
charge $4$ compatible with the metric in Eq.~(\ref{simplem}).

Another interesting quantity is the ratio
\begin{equation*}
\Delta =\frac{\text{Vol}\left( SU(3)\right) }{\text{Vol}\left( SU(2)\right) }
=\frac{1}{2\sqrt{2}}\left(\frac{2K\kappa -1}{K\kappa-2}\right)^{\frac{3}{2}}\ ,
\end{equation*}
between the three-dimensional volume of the spatial section of the static
gravitating $4$-Baryons (corresponding to the static solutions of Eq.
(\ref{isosk})) and the static gravitating Skyrmions of \cite{canfora6}
(corresponding to the static solutions of Eq.~(\ref{isoskSU(2)})). It reveals
the non-trivial strong and gravitational interactions of the system (as, in
the case of four non-interacting solitons, one should expect $\Delta =4$).

The above regular solutions of the Einstein-$\Lambda$-$SU(3)$-Skyrme system
defined in Eqs.~(\ref{simplem}), (\ref{range1}), (\ref{bala}), (\ref{bala2})
and (\ref{skbala}), are the first analytic self-gravitating Skyrmions of
higher Baryonic charge in $(3+1)$-dimensions.

The fact that the Baryon charge is $4$ instead of $2$ as in
\cite{bala0,Bala1}, is related to the compactness of the $t=\text{const.}$
hypersurfaces of the metric (\ref{simplem}). Here, rather than requiring
boundary conditions at spatial infinity as in \cite{bala0,Bala1}, one has to
require periodic boundary conditions for $U_{\text{B}}$ compatible with the
compact spatial metric (\ref{simplem})-(\ref{range1}). This charge $4$ arises
due to the fact that, unlike what happens in \cite{bala0,Bala1}, the present
gravitating $4$-Baryons must wrap around three compact spatial directions
instead of two.

\subsection{Bianchi IX}

The above construction can be further extended to a Bianchi IX cosmology.
Indeed, the $SU(3)$ Skyrme field equations on the metric
\begin{align}
ds^{2}& =-dt^{2}+I_{1}^{2}(\cos \theta d\gamma +d\phi )^{2}
+I_{2}^{2}(\cos \phi d\theta +\sin \theta \sin \phi d\gamma)^{2}
+I_{3}^{2}(\sin \phi d\theta -\sin \theta \cos \phi d\gamma )^{2}\ ,
\label{BianchiIX}
\end{align}
where $I_{j}=I_{j}(t)$, are still identically satisfied with the same ansatz
defined in Eqs.~(\ref{bala}), (\ref{bala2}) and (\ref{skbala})! The reason
behind this quite remarkable fact is that the left-invariant forms that
appear in the construction of the most general Bianchi IX metric are
proportional to the left-invariant forms that appear when computing the
Skyrmion derivatives
\begin{equation*}
\left( U_{\text{B}}\right) ^{-1}\partial _{\mu }U_{\text{B}}
=\Omega _{\mu }^{a}\lambda_{a}\ ,
\end{equation*}
with the $U_{\text{B}}$ defined in Eqs.~(\ref{bala}), (\ref{bala2}) and
(\ref{skbala}). Since the $\Omega _{\mu }^{a}$ characterizing the Skyrmionic
configuration play, at the same time, also the role of \textquotedblleft
drei-bein'' of the spatial metric, huge simplifications appear in the field
equations. Hence, the complete set of coupled Einstein-$\Lambda
$-$SU(3)$-Skyrme field equations (\ref{skyrme-einstein}) and (\ref{tmunu})
for the metric (\ref{BianchiIX}) with the ansatz in Eqs.~(\ref{bala}),
(\ref{bala2}) and (\ref{skbala}), reduce to a consistent dynamical system for
the three Bianchi IX scale factors
\begin{widetext}
\begin{align*}
I^{(4)}-2\mathcal{I}+4\Lambda \mathcal{I}_{(3)}^{2}-4\mathcal{I}_{(3)}\left(
I_{1}^{\prime }I_{2}^{\prime }I_{3}+I_{1}^{\prime }I_{2}I_{3}^{\prime
}+I_{1}I_{2}^{\prime }I_{3}^{\prime }\right) +\frac{K\kappa }{2}(\lambda
I^{(2)}+4\mathcal{I})& =0\ , \\
\left( 4I_{1}^{4}-I^{(4)}-2\mathcal{I}+4I_{2}^{2}I_{3}^{2}\right)
I_{2}I_{1}^{\prime }+\left( 4I_{1}^{4}-2I^{(4)}+4(1-\Lambda
I_{1}^{2})I_{2}^{2}I_{3}^{2}\right) I_{1}I_{2}^{\prime }& \\
{}+4\mathcal{I}_{(3)}I_{3}\left( I_{1}I_{1}^{\prime }I_{2}^{\prime
2}+I_{2}(I_{1}I_{2})^{\prime }I_{1}^{\prime \prime }\right)
+\frac{K\kappa}{2}\left[ \lambda I_{1}\left( 2I_{1}(I_{1}I_{2})^{\prime
}-I^{(2)}I_{2}^{\prime }\right) +4I_{2}\left( \mathcal{I}I_{1}^{\prime
}-2I_{2}I_{3}^{2}(I_{1}I_{2})^{\prime }\right) \right] & =0\ , \\
\end{align*}
and
\begin{align*}
\left( 4I_{2}^{4}-I^{(4)}-2\mathcal{I}+4I_{1}^{2}I_{3}^{2}\right)
I_{1}I_{2}^{\prime }+\left( 4I_{2}^{4}-2I^{(4)}+4(1-\Lambda
I_{2}^{2})I_{1}^{2}I_{3}^{2}\right) I_{2}I_{1}^{\prime }& \\
{}+4\mathcal{I}_{(3)}I_{3}\left( I_{2}I_{2}^{\prime }I_{1}^{\prime
2}+I_{1}(I_{1}I_{2})^{\prime }I_{2}^{\prime \prime }\right)
+\frac{K\kappa}{2}\left[ \lambda I_{2}\left( 2I_{2}(I_{1}I_{2})^{\prime
}-I^{(2)}I_{1}^{\prime }\right) +4I_{1}\left( \mathcal{I}I_{2}^{\prime
}-2I_{1}I_{3}^{2}(I_{1}I_{2})^{\prime }\right) \right] & =0\ ,
\end{align*}
\end{widetext}
where we have defined $I^{(2)}=I_{1}^{2}+I_{2}^{2}+I_{3}^{2}$,
$I^{(4)}=I_{1}^{4}+I_{2}^{4}+I_{3}^{4}$, $\mathcal{I}
=I_{1}^{2}I_{2}^{2}+I_{1}^{2}I_{3}^{2}+I_{2}^{2}I_{3}^{2}$,
$\mathcal{I}_{(3)}=I_{1}I_{2}I_{3}$.

As the $SU(3)$ Skyrme field equations are automatically satisfied, the above
dynamical system for the scale factors $I_{j}\left( t\right) $ describes the
dynamical evolution of the four self-gravitating Skyrmions. The analysis of
such system can reveal many interesting features about the interplay between
the gravitational and the strong interactions of these $4$-Baryons. A
comparison with the numerical gravitating Skyrmions constructed in
\cite{satosawado} is useful. Their numerical solutions with Baryon charge $2$
(which are based on the $SU(2)$ subgroup of $SU(3)$) are asymptotically flat
while the present analytic solutions with Baryon charge $4$ (which are based
on a non-embedded ansatz) either have compact spatial sections (in the
Bianchi IX sector) or possess two asymptotically NUT-AdS regions (see the
section below).

\section{NUT-AdS Wormhole}

In this section we will consider the limit of vanishing Skyrme coupling
$\lambda =0$. Following \cite{canfora6}, a double-Wick rotation of the ansatz
in Eq.~(\ref{skbala}) leads to $\psi =0$ and
\begin{equation}
\Phi =\frac{t+\varphi }{2},\quad \tan \Theta
=\frac{\cot \left( \frac{\theta }{2}\right) }
{\cos \left( \frac{t-\varphi }{2}\right) },\quad \tan \left( \frac{\chi
}{2}\right) =\frac{\sqrt{1+\tan ^{2}\Theta }}
{\tan \left( \frac{t-\varphi }{2}\right) },  \label{balawor}
\end{equation}
for the $SU(3)$ non-linear sigma model. Interestingly enough, the field
equations of the $SU(3)$ non-linear sigma model with the ansatz defined in
Eqs.~(\ref{bala}), (\ref{bala2}) and (\ref{balawor}) on the metric
\begin{equation}
ds^{2}=\rho (\gamma )^{2}\left[ -Q^{2}(dt+\cos {\theta }d\varphi
)^{2}+d\theta ^{2}+\sin ^{2}{\theta }d\varphi ^{2}\right] +d\gamma ^{2}\ ,
\label{balawormetric}
\end{equation}
are identically satisfied. A direct computation shows that the coupled
Einstein-$\Lambda $-$SU(3)$-non-linear sigma model field equations
(\ref{skyrme-einstein}) and (\ref{tmunu}) with $\lambda =0$ are satisfied for
\begin{equation*}
\rho(\gamma)=\sqrt{\frac{3(\kappa K-2)}{4|\Lambda|}}
\cosh\left(\sqrt{{\frac{|\Lambda|}{3}}}{\gamma}\right) \ ,\quad \
Q^{2}=\kappa K\ ,
\end{equation*}
provided $\Lambda <0$. Notice that the square of the NUT parameter $Q^{2}$ is
four times the result for $SU(2)$ \cite{canfora6}. The above metric
represents a traversable Lorentzian wormhole with NUT-AdS asymptotic
supported by a reasonable physical source (see \cite{canfora6} \cite{cawor}
); the non-linear sigma models do not violate energy conditions. The
well-known no-go results on the existence of wormholes are avoided due to the
presence of a NUT parameter. The $SU(3)$ wormhole throat is larger than the
one of the $SU(2)$ configuration in \cite{canfora6,cawor} since the present
matter field possesses a higher topological charge.

\section{$4$-Baryons at finite density}

The first analytic solutions with non-vanishing Baryon charge in the $SU(2)$
Skyrme model on flat space have been found by adapting the self-gravitating
solution found in \cite{canfora6} to a flat space with finite volume in
\cite{canfora9}. A similar construction also works in the present case. The
sector explored below describes four low-energy Baryons confined within a
finite volume in flat space.

Consider the $SU(3)$ Skyrme configuration defined by $\psi =0$ and
\begin{equation}
\Phi =\frac{\gamma +\varphi }{2},\quad \tan \Theta = \frac{\tan H\left(
t,z\right) }{\cos \left( \frac{\gamma -\varphi }{2} \right) },\quad
\tan \left(\frac{\chi }{2}\right) =\frac{\sqrt{1+\tan ^{2}\Theta }}
{\tan \left( \frac{\gamma -\varphi }{2}\right) },  \label{flatBox}
\end{equation}
together with Eqs.~(\ref{bala}) and (\ref{bala2}) in a flat metric of the
form
\begin{equation}
ds^{2}=-dt^{2}+L_0^2\left[ dz^{2}+d\gamma ^{2}+d\varphi ^{2}\right] \ ,
\label{flatmetric}
\end{equation}
where $L_{0}$ has dimension of length and represents the size of the box in
which the Baryons are confined, and the dimensionless coordinates have the
following ranges
\begin{equation*}
0\leq z\leq 2\pi \ ,\quad 0\leq \gamma \leq 4\pi \ ,\quad 0\leq \varphi \leq
2\pi \ .
\end{equation*}

The topological density is again different from the $SU(2)$ case since
\begin{equation*}
\rho _{\text{B}}^{\text{SU(3)}}=-12\sin (2H)\partial _{z}H\ ,\qquad
\rho _{\text{B}}^{\text{SU(2)}}=-3\sin (2H)\partial _{z}H\ .
\end{equation*}
Considering the boundary conditions as $H(t,0)=0$ and
$H(t,2\pi)=\pm\frac{\pi}{2}$, the $SU(3)$ topological charge is $\pm 4$
(which is four times the charge found in
\cite{canfora9,canfora8,Canfora:2018clt}). Concretely, these Skyrmions
confined to a finite volume can only have charges $-4$, $0$, or $4$ (while in
the $SU(2)$ case defined in \cite{canfora9,canfora8,Canfora:2018clt}, the
corresponding Skyrmions can only have charges $-1$, $0$, or $1$). This
confirms the genuine $SU(3)$ nature of the ansatz defined in
Eqs.~(\ref{bala}), (\ref{bala2}) and (\ref{flatBox}).

The $SU(3)$ Skyrme field equations on the flat metric (\ref{flatmetric}) for
the Baryon-like ansatz (\ref{bala}), (\ref{bala2}) and (\ref{flatBox}) become
a single partial differential equation for a scalar a profile
\begin{equation}\label{sineflat}
\Box H-\frac{\lambda }{8L_{0}^{2}(2L_{0}^{2}+\lambda )}\sin (4H)=0\ .
\end{equation}
Hence, the eight coupled $SU(3)$ Skyrme field equations collapse to a single
integrable PDE. This opens the intriguing possibility to analyze many
interesting non-trivial properties of these multi-Skyrmions confined to a
finite volume using Sine-Gordon theory such as possible phase transitions
between the trivial embedding of the $SU(2)$ solutions \cite{canfora9} and
the present non-embedded $SU(3)$ solutions living in the same box.

At this point it is important to recall that the value $4$ of the Baryonic
charge cames only from the nature of the $SU(3)$ ansatz and the compactness
of the space-time under consideration, both, in the present case and in the
previous gravitating configurations. Using the same boundary conditions for
the usual $SU(2)$ Skyrme field is not possible to obtain $B=4$. In fact, Eqs.
\eqref{flatBox} and \eqref{sineflat} describes a bound state of $4$-Baryons
whose energy density is localized. In the next section will be show that
indeed, the energy of the $4$-Baryon state is different from the one of the
four Skyrmions on the $SU(2)$ ansatz.

\section{A novel transition at finite Baryon density}

In the present section we will show that, within the same box of finite
volume, there is a competition between embedded and non-embedded
configurations with the same Baryonic charge, the control parameter of the
transition being the Baryon density. We show here below that at high Baryon
density non-embedded solutions are favoured over embedded solutions (and
vice-versa at low Baryon density).

A suitable ansatz to describe $SU(3)$ configurations living in the flat box
defined above that are trivial embedding of $SU(2)$ into $SU(3)$ is
\begin{eqnarray}
\vec{\Lambda} &=&(\lambda _{1},\lambda _{2},\lambda _{3})\ ,\quad
\Phi =\frac{p\gamma +q\varphi }{2}\ ,  \notag \\
\quad \tan \Theta &=&\frac{\tan H\left( t,z\right) }
{\cos \left( \frac{p\gamma -q\varphi }{2}\right) }\ ,\quad
\tan \left( \chi \right) =\frac{\sqrt{1+\tan ^{2}\Theta }}
{\tan \left( \frac{p\gamma -q\varphi }{2}\right) }\ ,\label{flatBox2}
\end{eqnarray}
where $p$ and $q$ are non-vanishing integer numbers (see \cite{last}) and
$\lambda_{1}$, $\lambda_{2}$, $\lambda_{3}$ the first three Gell-Mann
matrices.

As we explained in Section V, as well as in \cite{last}, both in the
$4$-Baryon case (Eqs. (\ref{bala}) and (\ref{flatBox})) and in the case of
trivially embedded solutions (Eqs. (\ref{bala}) and (\ref{flatBox2})) the
consistent boundary conditions for the profile $H(t,r)$ are
\begin{equation*}
H(t,0)=0\ ,\ \ H(t,2\pi )=\pm \frac{\pi }{2}\ .
\end{equation*}

The topological densities read
\begin{equation*}
\rho _{\text{B}}^{\text{SU(3)}}=-12\sin (2H)\partial _{z}H\ ,\qquad
\rho _{\text{B}}^{\text{SU(2)}}=-3pq\sin (2H)\partial _{z}H\ .
\end{equation*}
We know that the topological charge in the $SU(3)$ case is
$B^{\text{SU(3)}}=4$, while the topological charge in the trivially embedded
solutions with the above slightly generalized ansatz is
\begin{equation*}
B=pq\ ,
\end{equation*}
where the integers $p$ and $q$ are the ones appearing in the ansatz in
Eqs.~(\ref{bala}) and (\ref{flatBox2}) (see also \cite{last}).

Thus, if we choose
\begin{equation*}
pq=4\ ,
\end{equation*}
then we will have two inequivalent configurations\footnote{Using the elegant
arguments in \cite{bala0}, one can easily see that these two configurations
cannot be deformed into each other using global isospin transformations.}
(the one in Eqs.~(\ref{bala}) and (\ref{flatBox}) as well as the one in
Eqs.~(\ref{bala}) and (\ref{flatBox2})) with the same charge living in the
same box. The natural question is:

\textit{Which type of configuration is energetically favoured?}

Thanks to the remarkable properties of the ansatz described above, one can
answer this question explicitly. The first step is to notice that in both
cases (the one in Eqs.~(\ref{bala}) and (\ref{flatBox}) as well as the one in
Eqs.~(\ref{bala}) and (\ref{flatBox2})) the $SU(3)$ Skyrme field equations in
the flat metric (\ref{flatmetric}) reduce to a single partial differential
equation of the form
\begin{equation}
\Box H-\beta \sin (4H)=0\ ,  \label{SineGordon}
\end{equation}
with
\begin{equation*}
\beta _{\text{SU(3)}}=\frac{\lambda }{8L_{0}^{2}(2L_{0}^{2}+\lambda )}\
,\quad
\beta _{\text{SU(2)}}=\frac{p^{2}q^{2}\lambda }
{4L_{0}^{2}(4L_{0}^{2}+\lambda (p^{2}+q^{2}))}\ ,
\end{equation*}
for (\ref{flatBox}) and (\ref{flatBox2}), respectively.

For static configurations $H(t,r)=H(r)$, Eq.~(\ref{SineGordon}) admits a
first integral
\begin{equation*}
\left( H^{\prime }\right) ^{2}+\frac{1}{2}\beta \cos (4H)=I_{0}\ ,
\end{equation*}
with the integration constant defined using the boundary conditions as
\begin{equation*}
\int_{0}^{2\pi }dr=\int_{0}^{\frac{\pi }{2}}\frac{dH}{\sqrt{I_{0}-\frac{1}{2}
\beta \cos (4H)}}=2\pi \ .
\end{equation*}
The energy density for these configurations are given by
\begin{align*}
T_{00}^{\text{SU(3)}}& =-\frac{K}{16L_{0}^{4}}\left( 16L_{0}^{2}+\lambda
(1-\cos (4H))+\frac{16}{L^{2}}(2L_{0}^{2}+\lambda )H^{\prime 2}\right) \ , \\
T_{00}^{\text{SU(2)}}& =-\frac{K}{16L_{0}^{4}}\biggl(
8L_{0}^{2}(p^{2}+q^{2})+p^{2}q^{2}\lambda (1-\cos
(4H))+8(4L_{0}^{2}+(p^{2}+q^{2})\lambda )H^{\prime 2}\biggl)\ .
\end{align*}%
\begin{table}[tbp]
\begin{center}
\begin{tabular}{||c||c|c|c|c||}
\hline
$L_0$ & $E^{SU(2)}_{2,2}$ & $E^{SU(2)}_{1,4}$ & $E^{SU(2)}_{4,1}$ &
$E^{SU(3)}$ \\[0.5ex] \hline\hline
0.01 & 283.131 & 416.22 & 416.22 & 147.81 \\
0.1 & 29.627 & 44.317 & 44.317 & 16.198 \\
0.5 & 12.288 & 21.925 & 21.925 & 10.082 \\
1 & 16.040 & 31.359 & 31.359 & 15.675 \\
1.4 & 20.455 & 41.033 & 41.033 & 20.899 \\
2 & 27.637 & 56.414 & 56.414 & 29.054 \\
3 & 40.093 & 82.814 & 82.814 & 42.933 \\
5 & 65.557 & 136.421 & 136.421 & 70.999 \\
10 & 129.980 & 271.440 & 271.440 & 141.528 \\ \hline
\end{tabular}
\end{center}
\caption{Phase transition $SU(3)$ vs. $SU(2)\subset SU(3)$ at finite
density.}
\label{T1}
\end{table}

TABLE \ref{T1}\footnote{The notation $E_{X,Y}^{SU(2)}$ means ``the energy of
the $SU(2)$ configuration in Eqs. (\ref{bala}) and (\ref{flatBox2}) with
$p=X$ and $q=Y$''. In the table $XY=4$ as we need to compare configurations
with the same topological charge.} shows the energy comparison between the
$SU(2)$ and the $SU(3)$ configurations with Baryon charge $B=4$, while it
varies the size of the box (measured by $L_{0}$). We have fixed $K=2$,
$\lambda =1$ (which means that we are measuring length in $fm$). One can see
that a phase transition appears at\footnote{It is worth to mention that at
this scale, the Skyrme model is still well within its range of validity.}
$L_{0}^{\ast }\approx 1.4$. When $L_{0}<L_{0}^{\ast }$ then the $4$-Baryon
are favoured over the trivially embedded solutions (and vice-versa when
$L_{0}>L_{0}^{\ast }$). The appearance of this transition is related to the
fact that $4$-Baryons deal more efficiently with the well known repulsive
interaction between Skyrmions. From the plots of the energy densities below
for the two types of configurations one can see that, at high Baryon density,
it pays off that the $4$-Baryon is ``less peaked'' around its maximum while
at low Baryon densities the energy densities of both types of solutions
become relatively flat and, in such cases, the trivially embedded solutions
prevail.%
\begin{figure}[h]
\centering
\includegraphics[scale=0.45]{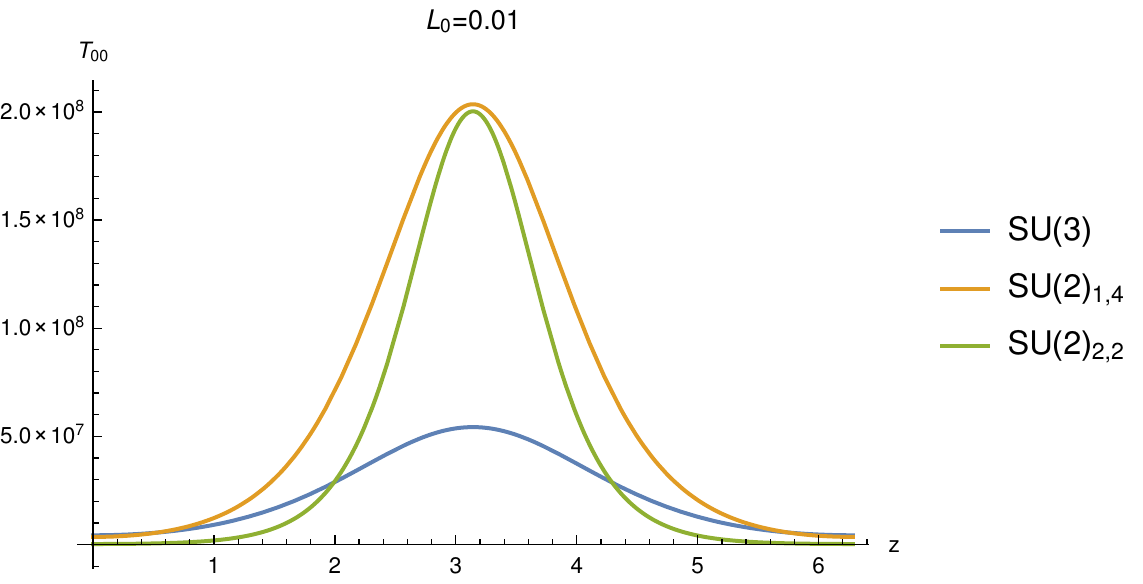}\qquad
\includegraphics[scale=0.45]{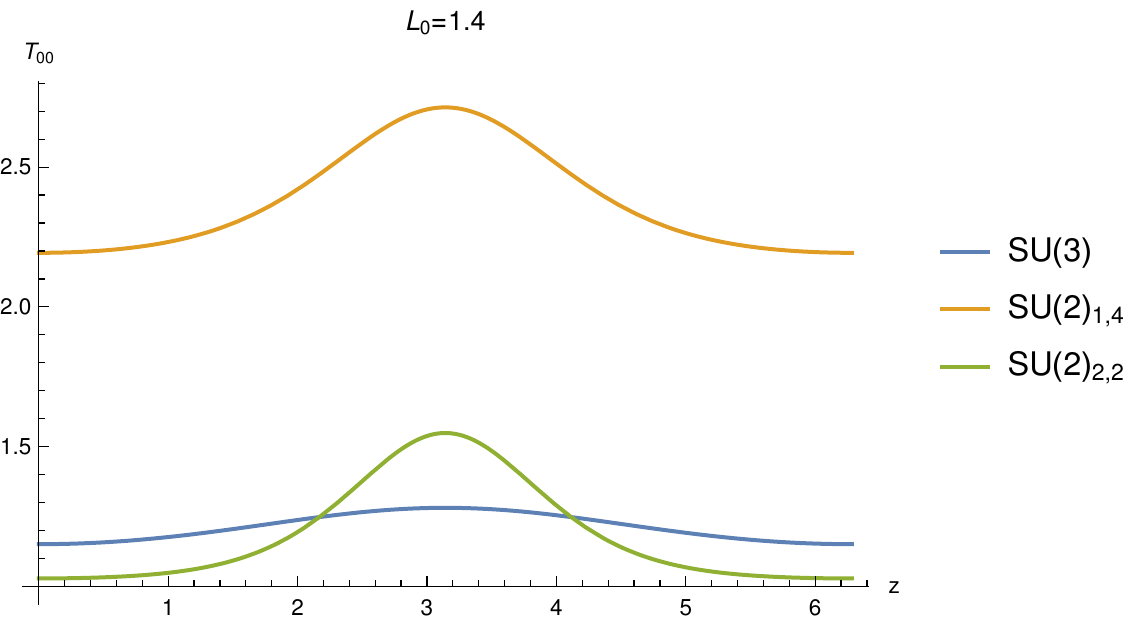}\qquad
\includegraphics[scale=0.45]{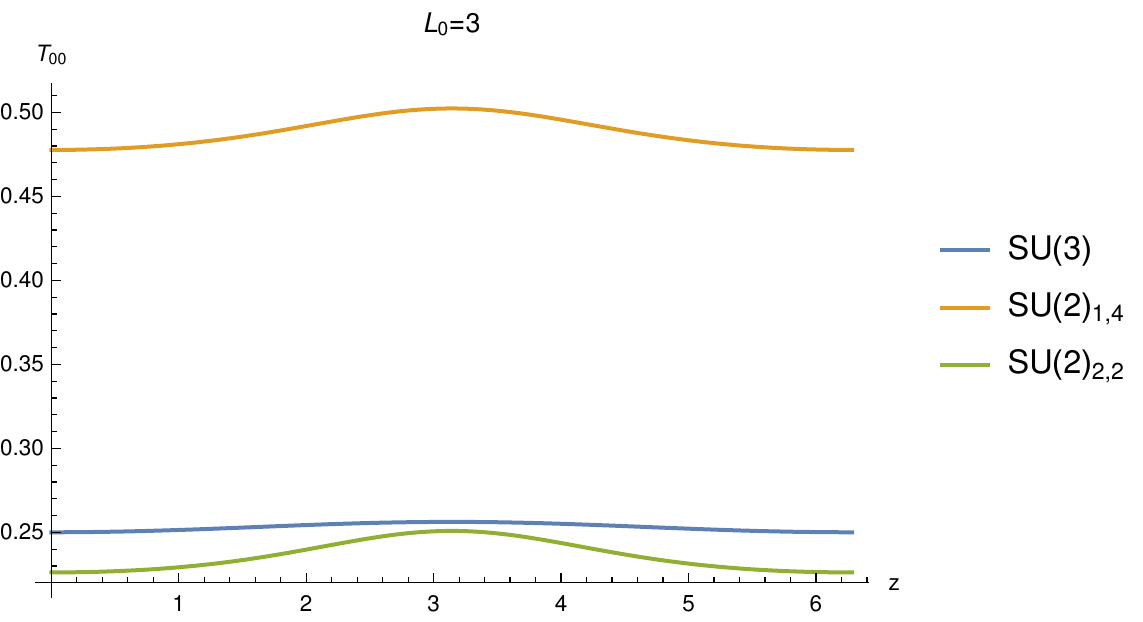}
\caption{Energy densities for different values of $L_{0}$. We can see that
at high Baryon density it is convenient that the $4$-Baryon is
``less localized'' around its maximum, and at low Baryon density the energy
density for both types of solutions tend to flatten.}
\end{figure}

Fig.~1 shows the energy density of the three relevant configurations, for
three different values of the size of the box.

At last, we make a small observation about the quantization of the solutions.
Since, with the above ansatz, the hedgehog property holds (as the $8$ coupled
$SU(3)$ field equations reduce to a single PDE for the profile $H$). The
small fluctuations of the profile $H$ around these solutions are described by
the effective action obtained replacing the ansatz itself into the original
$SU(3)$ Skyrme action (\cite{coleman}, \cite{shifman}). Thus, the
quantization of the $4$-Baryon living in the finite box defined above can be
analyzed using known results on the sine-Gordon theory plus the semiclassical
quantization of the Isospin degrees of freedom described in \cite{Bala1}.
This observation also shows that the flat configurations constructed in the
manuscript are stable, at least, under the perturbations which keep the
hedgehog properties. Namely, the flat solutions considered here are stable
under the following type of perturbations (see the analysis in \cite{coleman}
and \cite{shifman}):
\begin{equation*}
H(t,r)\rightarrow H(t,r)+\varepsilon u\left( t,r\right) \ ,\ \ 0<\varepsilon
\ll 1\ .
\end{equation*}
We hope to come back on the interesting but rather difficult problem of the
full stability analysis (which must be analyzed numerically) in a future
publication.

\section{Conclusions and perspectives}

The first regular analytic self-gravitating Skyrmions of higher Baryon charge
have been constructed. The space-time corresponds to a general Bianchi IX
cosmology whose three scales factors evolve accordingly, while the Skyrme
field equations are identically satisfied in the sector with Baryon charge
$4$. All these solutions disclose genuine features of the $SU(3)$ Skyrme
model. Traversable wormholes with NUT-AdS asymptotic supported by an $SU(3)$
regular solitonic solution of the resulting non-linear sigma model are also
constructed. Also, a suitable flat limit in a finite volume of the
self-gravitating Skyrme configurations can also be defined. In this case, the
eight $SU(3)$ Skyrme field equations become the Sine-Gordon field theory
without sacrificing the higher Baryon charge. This flat limit explicitly
describes charge $4$ Baryons confined in a box. The present formalism
discloses the existence of a novel transition between embedded and
non-embedded configurations with the same Baryonic charge. The control
parameter of this transition is the Baryon density: at high Baryon densities,
it is energetically convenient to have non-embedded solutions while, at low
Baryon densities, trivially embedded solutions prevail. To the best of
authors knowledge, this phenomenon related to the competition between
embedded and non-embedded solutions at finite Baryon density is new. The
reason is that the present techniques are especially suitable to deal with
Skyrmions and $4$-Baryons at finite density as these techniques allow to
determine how relevant physical properties of these solitons depend on the
Baryon density. Without these informations it would have been impossible to
disclose such possibility.

\begin{acknowledgments}
M.L. and A.V. appreciate the support of FONDECYT postdoctoral grants 3190873
and 3200884. This work has been partially funded by the Fondecyt grants
1200022, 1181047, and the Conacyt grant A1-S-11548. The Centro de Estudios
Cient\'{\i}ficos (CECs) is funded by the Chilean Government through the
Centers of Excellence Base Financing Program of Conicyt.
\end{acknowledgments}

\end{document}